\documentclass[prmaterials,psfig,pstricks,twocolumn,preprintnumbers,amsmath,amssymb,superscriptaddress]{revtex4-1}

\usepackage{graphicx}          
\usepackage{dcolumn}           
\usepackage{bm}                

\usepackage{array,booktabs,lmodern}
\usepackage[scaled=1.00]{helvet}
\usepackage[final]{listings,microtype}
\usepackage[T1]{fontenc}
\usepackage[version=3]{mhchem} 
\usepackage{siunitx} 
\usepackage[english]{babel}
\usepackage{float}
\usepackage{natbib}
\usepackage{setspace}
\usepackage{xkeyval}
\usepackage{natmove}
\usepackage{titlesec}
\usepackage{gensymb}

\newcommand{\mat}[1]{\mbox{\boldmath{$#1$}}} 
\usepackage{xfrac}
\usepackage{relsize}
\newcommand{\ie}{\textit{i}.\textit{e}.}

\begin{document}


\title{Computational prediction of high thermoelectric performance in As$_{2}$Se$_{3}$ by engineering out-of-equilibrium defects}

\author{Anderson S. Chaves}
\email{aschaves@ifi.unicamp.br}
\affiliation{John A. Paulson School of Engineering and Applied Sciences, Harvard University, Cambridge, Massachusetts 02138, USA}
\author{Murilo Aguiar Silva}
\email{aguiarsm@ifi.unicamp.br}
\affiliation{Instituto de F\'{i}sica Gleb Wataghin, Universidade Estadual de Campinas, UNICAMP, 13083-859 Campinas, S\~{a}o Paulo, Brazil}
\author{Alex Antonelli}
\email{aantone@ifi.unicamp.br}
\affiliation{Instituto de F\'{i}sica Gleb Wataghin and Centre for Computational Engineering \& Sciences, Universidade Estadual de Campinas, UNICAMP, 13083-859 Campinas, S\~{a}o Paulo, Brazil}

\date{\today}

\begin{abstract}

We employed first-principles calculations to investigate the thermoelectric transport properties of the compound As$_2$Se$_3$. Early experiments and calculations
have indicated that these properties are controlled by a kind of native defect called antisites. Our calculations using the linearized Boltzmann transport equation within the relaxation time approximation show good agreement with the experiments for defect concentrations of the order of 10$^{19}$ cm$^{-3}$. Based on our total energy 
calculations, we estimated the equilibrium concentration of antisite defects to be about 10$^{14}$ cm$^{-3}$. These results suggest that 
the large concentration of defects in the experiments is due to kinetic and/or off-stoichiometry effects and in principle it 
could be lowered, yielding relaxation times similar to those found in other chalcogenide compounds. In this case, for relaxation
time higher than 10 fs, we obtained high thermoelectric figures of merit of 3 for the p-type material and 2 for the n-type one. 

\end{abstract}


\maketitle

\section{Introduction}
The plethora of environmental issues society faces nowadays has driven the search for more efficient energy  sources. To this end, the direct conversion of waste heat from engines and industrial processes into electricity through the use of thermoelectric generators has motivated a great deal of scientific research on thermoelectric materials \cite{Beretta2019}. Since the discovery of the outstanding thermoelectric performance of SnSe \cite{Zhao2014}, there has been an increasing interest in the study of the thermoelectric transport properties of chalcogenide materials, in particular binary compounds, such as SnSe. Therefore, the search for feasible candidates among binary chalcogenide compounds has increased in the last few years \cite{Jia2020}. 

The thermoelectric performance of a material is measured by its figure of merit, which is a dimensionless quantity, $zT$, given by the equation, 
\begin{equation}
	zT = \frac{\sigma S^{2} T}{\kappa_{e}+\kappa_{L}},
\end{equation}
where $\sigma$ is the electrical conductivity, $S$ is the Seebeck coefficient, $T$ is the temperature, $\kappa_{e}$ is the carrier thermal conductivity, and $\kappa_{L}$ is the lattice, or phonon, thermal conductivity. SnSe owes its remarkable figure of merit to its ultralow lattice thermal conductivity \cite{Zhao2014}. Few years ago, it was found, through first-principles calculations, that another layered binary chalcogenide compound As$_2$Se$_3$ also exhibits ultralow phonon thermal conductivity, as low as 0.14 W m$^{-1}$ K$^{-1}$ along the direction perpendicular to the layers \cite{Gonzalez-Romero2018}. This feature of both materials may be due to the repulsion between the lone-pair electrons of groups IV and V  cations and the p orbitals electrons of the chalcogenide anions, which enhances the anharmonicity of the chemical bonds \cite{Nielsen2013}. As$_2$Se$_3$ has been extensively investigated, mainly its glass phase, in the 1970's and 80's because of its  optical properties \cite{Tanaka2021}. However, very little has been done regarding the transport properties of its crystalline phase \cite{Wood1976,Marshall1977,Brunst1985}. Thus, the investigation of the thermoelectric transport properties of crystalline As$_2$Se$_3$ is required.

\begin{figure}
	\centering
	\includegraphics[width=0.45\textwidth]{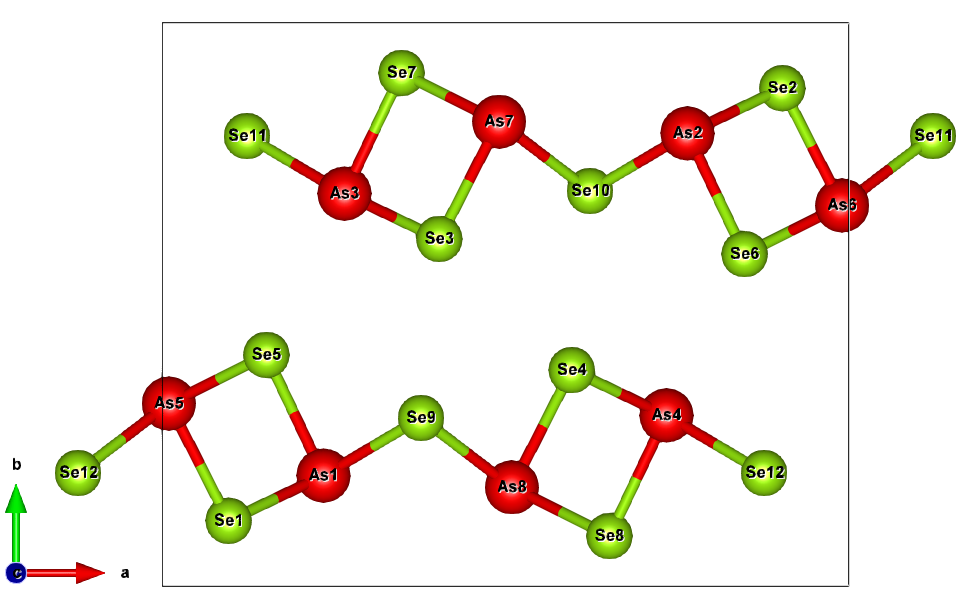}
	\vspace{-3mm}
	\caption{Crystalline structure of As$_2$Se$_3$. As atoms depicted in green and Se atoms shown in red.}
	\label{structure}
\end{figure}

Crystalline As$_2$Se$_3$ has monoclinic symmetry. A view of the unit cell along the c-axis is shown in Fig.~\ref{structure}, which comprises a double layer, containing 20 atoms altogether. The two layers are identical, however they are rotated by 180$\degree$ with respect to each other. The b-axis is perpendicular to the layers and the calculated angle between the c-axis and the a-axis is 90.458$\degree$ \cite{Stergiou1985}. The layers can be seen as helical chains, composed by alternating Se and As atoms, along the c-axis, which are linked by Se atoms. The projection of the helical chains on the plane of Fig.~\ref{structure}  are seen as rectangles. Within the layers, bonding is covalent, while between layers it is typically van der Waals. As$_2$Se$_3$ is a semiconductor with a band gap of 2.0 eV \cite{Althaus1980}.

The measurements of electrical conductivity \cite{Wood1976,Marshall1977,Brunst1985} exhibit an Arrhenius behavior, indicating that carrier transport is a thermally activated process. The activation energies obtained from the experimental results are similar, however, the measured values of the electrical conductivity showed large variation among the experiments. The measured Seebeck coefficient \cite{Brunst1985} is negative, indicating that electrons are responsible for electrical transport. These results suggest that trapped electrons in mid-gap levels created by defects are involved in the electrical transport. It has been suggested that antisite defects, i.e., threefold coordinated Se$^+$ ions at As sites and twofold coordinated As$^-$ ions at Se sites, are the defects that create such mid-gap levels. It has been argued that these defects have lower formation energies than vacancies and interstitials, since no bond breaking occurs upon their formation \cite{Tarnow1989}. These defects can be thermally created, that means that, even under strict stoichiometric conditions, at finite temperature there would be an equilibrium concentration of these defects, which are created in pairs. However, deviations from stoichiometric conditions can also create such defects. Another feature of the material caused by these defects is the pinning of the Fermi level \cite{Brunst1985}. All the experiments show a typical Arrhenius behavior, \ie, a linear increase of the logarithm of the electrical conductivity with the inverse of temperature, strongly suggesting that electron scattering by ionized defects dominates transport relaxation time and electron-phonon scattering is irrelevant.

The ultralow lattice thermal conductivity of As$_2$Se$_3$ renders it a prospective material for thermoelectric applications, however, several questions should be addressed before it can be considered a promising candidate. The goal of this work is to shed light on some of these questions. For instance, what are the concentrations of antisite defects in the samples studied in the experiments? Are these equilibrium concentrations? If defect concentration could be reduced, thus, depinning the Fermi level, what would be the thermoelectric transport properties of the material?  

The approach used is a combination of first-principles calculations with the semiclassical Boltzmann transport equation (BTE) within the relaxation time approximation (RTA). This methodology has proven to be effective because it goes beyond parabolic or Kane models, and includes non-parabolicity, multiplicity and degeneracy of the band edges on the same footing \cite{Chaves2021}.

\section{Methodology}

\subsubsection{Theoretical Approach to the Transport Properties}

The transport properties of charge carriers in the diffusive regime are determined by solving numerically the
semiclassical BTE for the nonequilibrium 
carrier distribution function $f_{n,{\bf{k}}} = f(\epsilon_{n,{\bf{k}}})$, 
\begin{equation}
\label{boltz1}
\frac{\partial f_{n,{\bf{k}}}}{\partial t} + {\bf{v}}_{n,{\bf{k}}}\cdot\nabla_{{\bf{r}}} f_{n,{\bf{k}}} - \frac{{\bf{F}}}{\hbar}\cdot\nabla_{{\bf{k}}} f_{n,{\bf{k}}} = \left(\frac{\partial f_{n,{\bf{k}}}}{\partial t}\right)_{coll}~, 
\end{equation}
in which, $n$ and ${\bf{k}}$ are the band index and the wavevector, respectively. The
carrier band velocity in the $\{n,{\bf{k}}\}$ state with energy $\epsilon_{n,{\bf{k}}}$ 
is given by ${\bf{v}}_{n,{\bf{k}}} = 1 / {\hbar} \nabla_{{\bf{k}}}\epsilon_{n,{\bf{k}}}$,
and ${\bf{F}}$ is the Lorentz force acting upon the carriers.
The right-hand side of the equation describes the charge carrier collisions
involved in the various scattering processes that
drive the system toward a steady state. The result of these scattering processes can be expressed
by the per-unit-time transition probability, $W(n,{\bf{k}}|j,{\bf{{k}^{\prime}}})$, of a carrier 
starting at the state $\{n,{\bf{k}}\}$ and ending at
state $\{j,{\bf{{k}^{\prime}}}\}$.
If a steady state is to be reached, the principle of detailed balance should apply, thus the number of charge carriers
reaching state $\{j,{\bf{{k}^{\prime}}}\}$ from $\{n,{\bf{k}}\}$ should be the same as
the number departing state $\{j,{\bf{{k}^{\prime}}}\}$ toward state $\{n,{\bf{k}}\}$, in such a way that
\begin{multline}
\label{boltz2}
\left(\frac{\partial f_{n,{\bf{k}}}}{\partial t}\right)_{coll} = \sum_{j,{\bf{{k}^{\prime}}}} [W(j,{\bf{{k}^{\prime}}} | n,{\bf{k}}) f_{j,{\bf{{k}^{\prime}}}} \left(1 - f_{n,{\bf{k}}}\right) \\
- W(n,{\bf{k}} | j,{\bf{{k}^{\prime}}}) f_{n,{\bf{k}}} \left(1 - f_{j,{\bf{{k}^{\prime}}}}\right)]~.
\end{multline}
Once $f_{n,{\bf{k}}}$ is computed, all the transport properties can be readily evaluated. 

If the system is sufficiently close to equilibrium,
the nonequilibrium distribution function, $f_{n,{\bf{k}}}$, differs only slightly
from that of the equilibrium state, $f_{n,{\bf{k}}}^{(0)}$, in such a way that
$\Delta f(n,{\bf{k}}) = \lvert{f_{n,{\bf{k}}}-f_{n,{\bf{k}}}^{(0)}}\lvert << f_{n,{\bf{k}}}^{(0)}$. 
In this case, Eq.~\eqref{boltz2} can be written within the so-called RTA
\begin{equation}
\label{coll}
\left(\frac{\partial f_{n,{\bf{k}}}}{\partial t}\right)_{coll} = -\frac{\Delta f(n,{\bf{k}})}{\tau_{n,{\bf{k}}}}~, 
\end{equation}
where
\begin{multline}
\label{tau1}
\frac{1}{{\tau_{n,{\bf{k}}}}} = \sum_{{\bf{{k}^{\prime}}}} \sum_j W(n,{\bf{k}}|j,{\bf{{k}^{\prime}}}) \\
\left( \frac{1-f^{(0)}_{j,{\bf{{k}^{\prime}}}}}{1-f^{(0)}_{n,{\bf{k}}}} - \frac{f^{(0)}_{n,{\bf{k}}}}{f^{(0)}_{j,{\bf{{k}^{\prime}}}}}\frac{\Delta f(j,{\bf{{k}^{\prime}}})}{\Delta f(n,{\bf{k}})}\right)~, 
\end{multline}
in the case where quantum effects can be neglected, $W(n,{\bf{k}}|j,{\bf{{k}^{\prime}}})$ is independent of the magnetic field, ${\bf{B}}$ \cite{Askerov2010}.
Also, one can reasonably consider that $W(n,{\bf{k}}|j,{\bf{{k}^{\prime}}})$ does not depend on the electric field, ${\bf{E}}$, and the temperature gradient, $\nabla T$.
As a result, for a homogeneous system, in zero magnetic field, and a time-independent
electric field, in the steady-state regime, Eq.~\eqref{boltz1} becomes
\begin{equation}
\label{boltz3}
{\bf{v}}_{n,{\bf{k}}}\cdot\nabla_{{\bf{r}}} f_{n,{\bf{k}}} - \frac{e {\bf{E}}}{\hbar}\cdot\nabla_{{\bf{k}}} f_{n,{\bf{k}}} = -\frac{\Delta f(n,{\bf{k}})}{\tau_{n,{\bf{k}}}}~, 
\end{equation}
from which it is possible to determine the nonequilibrium 
distribution function, as long as $\tau_{n,{\bf{k}}}$ does not depend on 
${\bf{E}}$ or $\nabla T$. 

Moreover, since we regard the
deviation from equilibrium to be small, one can expand $f_{n,{\bf{k}}}$ to first order as 
\begin{equation}
\label{smallness}
f_{n,{\bf{k}}} = f_{n,{\bf{k}}}^{(0)} -{\tau}_{n,{\bf{k}}}{\bf{v}}_{n,{\bf{k}}}\cdot{\bf{\Phi_0}}(\epsilon)\left(\frac{\partial f^{(0)}}{\partial \epsilon}\right)~,
\end{equation}
in which ${\bf{\Phi_0}}(\epsilon) = -e{\bf{E^\prime}} - \frac{\epsilon - \mu}{T}\nabla T$ is 
the generalized disturbing force (dynamic and static) driving the system out
equilibrium, where 
${\bf{E^\prime}} = {\bf{E}} + (1/e)\nabla \mu = -\nabla (\phi_0-(\mu/e))$
is the gradient of the electrochemical potential. 
Using Eqs.~\eqref{smallness} and \eqref{tau1}, one obtains
\begin{multline}
\label{tau2}
\frac{1}{\tau_{n,{\bf{k}}}} = \sum_{{\bf{{k}^{\prime}}}} \sum_j W(n,{\bf{k}}|j,{\bf{{k}^{\prime}}}) \\
\frac{1-f^{(0)}(\epsilon^{\prime})}{1-f^{(0)}(\epsilon)}\left( 1 - \frac{{\tau}_{j,{\bf{{k}^{\prime}}}}}{\tau_{n,{\bf{k}}}} \frac{{\bf{v}}_{j,{\bf{{k}^{\prime}}}}\cdot\bf{\Phi_0}(\epsilon^{\prime})}{{\bf{v}}_{n,{\bf{k}}}\cdot\bf{\Phi_0}(\epsilon)}\right)~.
\end{multline}
If one considers that the per-unit-time probability of charge carrier transition, $W(n,{\bf{k}}|j,{\bf{{k}^{\prime}}})$, 
is a function only of the magnitudes of the vectors, $\lvert{\bf{k}}\lvert$ and
$\lvert{\bf{{k}^{\prime}}}\lvert$, and the angle between them, {${\bf{k}}\cdot{\bf{{k}^{\prime}}}$}, \ie, 
$W(n,{\bf{k}}|j,{\bf{{k}^{\prime}}}) = W_{n,j}(\lvert{\bf{k}}\lvert,\lvert{\bf{{k}^{\prime}}}\lvert,{\bf{k}}\cdot{\bf{{k}^{\prime}}})$.
Furthermore, regarding the dispersion relation as an arbitrary 
spherically symmetric function of the magnitude of the 
wavevector, $k = \lvert{\bf{k}}\lvert$,(not necessarily a parabolic function) and also that the charge carriers scattering  
is purely elastic, that is, 
with $\epsilon(\lvert{\bf{k}}\lvert)=\epsilon(\lvert{\bf{{k}^{\prime}}}\lvert)$ and consequently$\lvert{\bf{k}}\lvert = \lvert{\bf{{k}^{\prime}}}\lvert$, 
Eq.~\eqref{tau2}  may be further simplified as~\cite{Askerov2010}
\begin{equation}
\label{tau3}
\frac{1}{{\tau_{n,{k}}}} = \sum_{{\bf{{k}^{\prime}}}} \sum_j W(n,{\bf{k}}|j,{\bf{{k}^{\prime}}}) \left(1-\frac{{\bf{k}}\cdot{\bf{{k}^{\prime}}}}{k^2}\right)~.
\end{equation}
Here, some comments are in order. Although Eq.~\eqref{tau3} has been derived for isotropic bands, results obtained from it 
have been used to investigate transport properties of lead chalcogenides, which are anisotropic materials~\cite{Ahmad2010,Ravich1971}.
The usage of this methodology in this situation is possible due to the fact that, although anisotropic, the transport properties in these materials along the different directions are mutually independent, except for the case of magnetoresistance, which essentially depends on the anisotropy of the material. This methodology was applied in the case of SnSe yielding results in good agreement with the experiment for different values of temperature and chemical potential \cite{Chaves2021}.

One can now compute the charge carrier current density
\begin{equation}
	{\bf{j}} = -\frac{2e}{V}\sum_n\sum_{k}{\bf{v}}_{n,{\bf{k}}}f_{n,{\bf{k}}} = -\frac{2e}{(2\pi)^3}\sum_n\int{{\bf{v}}_{n,{\bf{k}}}f_{n,{\bf{k}}}d{\bf{k}}}~,
\end{equation}
and the heat energy flux density
\begin{multline}
	{\bf{j}}_Q = \frac{2}{V}\sum_n\sum_{{\bf{k}}}\left(\epsilon_{n,{\bf{k}}}-\mu\right){\bf{v}}_{n,{\bf{k}}}f_{n,{\bf{k}}} \\
	= \frac{2}{(2\pi)^3}\sum_n\int{\left(\epsilon_{n,{\bf{k}}}-\mu\right){\bf{v}}_{n,{\bf{k}}}f_{n,{\bf{k}}}d{\bf{k}}}~,
\end{multline}
where the factor $\num{2}$ is due to the electron
spin, $e$ stands for the electron charge, $V$ is the crystal's volume, and $\mu$ is the chemical potential.


Thermoelectric effects in anisotropic materials can be more conveniently studied using the tensorial formalism.
The derivation of the off-diagonal coupling between the electronic current density, ${\bf{j}}$, and
heat energy flux density, ${\bf{j}}_Q$, can be done
in a more general way using the linear response theory~\cite{Callen1985} as
\begin{equation}
	\begin{bmatrix} 
		{\bf{j}} \\ 
		{\bf{j}}_Q  
	\end{bmatrix}
	=
	\begin{bmatrix}
		{\bf{L^{11}}} & {\bf{L^{12}}} \\
		{\bf{L^{21}}} & {\bf{L^{22}}}
	\end{bmatrix}
	\cdot
	\begin{bmatrix} 
		{\bf{E^\prime}} \\
		-\frac{\nabla T}{T}
	\end{bmatrix}
\end{equation}
where, ${\bf{L^{11}}}$, ${\bf{L^{12}}}$, ${\bf{L^{21}}}$,
${\bf{L^{22}}}$ are the moments of the generalized transport
coefficients, which are called kinetic coefficient tensors, with
${\bf{L^{12}}} = {\bf{L^{21}}}$ according to the Onsager
reciprocity relations~\cite{Callen1985}.
These kinetic coefficients can be written as
\begin{equation}
	\label{Lambda}
	\Lambda^{(\alpha)}(\mu;T)= e{^2}\int\Xi(\epsilon,\mu,T)(\epsilon - \mu)^{\alpha}\left(-\frac{\partial f^{(0)}(\mu;\epsilon,T)}{\partial \epsilon}\right)d\epsilon~, 
\end{equation}
with ${\bf{L^{11}}} = \Lambda^{(0)}$,
${\bf{L^{21}}} = {\bf{L^{12}}} = -(1/e)\Lambda^{(1)}$, and
${\bf{L^{22}}} = (1/e^2)\Lambda^{(2)}$, in which $\Xi(\epsilon,\mu,T)$
is the transport distribution kernel (TDK) given by
\begin{equation}
	\Xi(\epsilon,\mu,T) = \int \sum_n{{\bf{v}}_{n,{\bf{k}}}\otimes{\bf{v}}_{n,{\bf{k}}}{\tau}_{n,{k}}}(\mu,T)\delta(\epsilon - \epsilon_{n,{\bf{k}}})\frac{d{\bf{k}}}{8\pi^3}~. 
	\label{TDK}
\end{equation}
Under the experimental condition of zero temperature gradient ($\nabla T = 0$),
the electrical conductivity tensor can be written as, $\sigma = \Lambda^{(0)}$, and under the experimental zero electric current condition,
the Seebeck coefficient tensor is given by, $S = (eT)^{-1}\Lambda^{(1)}(\Lambda^{(0)})^{-1}$,
while the charge carrier contribution to the thermal conductivity tensor can be identified as,
$\kappa_{e} = (e^2T)^{-1} \left({\Lambda^{(1)}\cdot(\Lambda^{(0)})^{-1}}\cdot{\Lambda^{(1)}} - \Lambda^{(2)}\right)$.


Now we turn to the calculation of the per-unit-time transition probability.
The so-called Born approximation in scattering theory considers that the magnitude of the Hamiltonian 
for charge carrier interaction, $H^{\prime}$ deviates 
only slightly from the magnitude of the non-perturbed Hamiltonian, $H$, that is, $H^{\prime} - H << H$. 
In this case, the transition probability per-unit-time between Bloch states $\Psi_{j,{\bf{{k}^{\prime}}}}$ and $\Psi_{n,{\bf{k}}}$ 
can be well described by the first order perturbation theory (Fermi's golden rule)
\begin{equation}
\label{1PT}
W(n,{\bf{k}}|j,{\bf{{k}^{\prime}}}) = \frac{2\pi}{\hbar}\lvert\langle\Psi_{j,{\bf{{k}^{\prime}}}}\lvert{H^{\prime}}\lvert\Psi_{n,{\bf{k}}}\rangle\lvert^2\delta(\epsilon_{j,{\bf{{k}^{\prime}}}}-\epsilon_{n,{\bf{k}}})~.
\end{equation}
The assumption is valid in the so-called weak coupling regime~\cite{Hameau1999}.
The total scattering rate is obtained by integrating out such probability over the Brillouin zone (BZ).
By using Eqs.~(\ref{1PT}) and~(\ref{tau3}), 
the expressions for the RTs associated with the different scattering 
mechanisms can be derived following Refs.~\cite{Chaves2021} and~\cite{Askerov2010}. 
From the previous analysis of the experimental results, here we are concerned only with the scattering by ionized defects.

In particular, we consider only the concentration of ionized defects
because the amount of charged donors or acceptors is usually considerably larger
than that of neutral imperfections. 
The scattering of carriers by ionized defect (impurity) has been studied theoretically by 
Brooks and Herring (B-H)~\cite{Brooks1955,Chattopadhyay1981}, by considering a screened Coulomb potential, the
Born approximation to calculate the transition probabilities,
and neglecting the perturbation effects of the defects on the electron energy levels and wave functions.
In the B-H approach, it is assumed that the carriers are scattered 
by dilute concentrations of ionized defects randomly distributed throughout the semiconductor.
It describes the scattering by charged defects without taking into account more complex effects, such as
contributions from coherent scattering by pairs
of ionized defects, which requires a quantum transport theory~\cite{Moore1967}.

One can write the per-unit-time transition probability for the scattering of charge carriers
by ionized defects in the plane-wave approximation as 
\begin{multline}
\label{U}
W({\bf{k}}|{\bf{{k}^{\prime}}}) = \frac{2\pi}{\hbar}\frac{N_i}{V} \\
\left\vert{\int U({\bf{r}}) exp\left[i({\bf{k}}-{\bf{{k}^{\prime}}})\cdot{\bf{r}}\right]d{\bf{r}}}\right\vert^2 \delta(\epsilon_{{\bf{{k}^{\prime}}}}-\epsilon_{{\bf{k}}})~,
\end{multline}
where $U({\bf{r}})$ is the scattering potential and $N_i$ is the ionized impurity concentration.

The long-range Coulomb potential energy,
$U(\mat{r}) = e\phi(r) = \pm e^2/\zeta_{0} r$, where $\phi$ is the potential at a point $r$ in the crystal, 
is due to the presence of positive (donor) or negative (acceptor)
ionized defects and $\zeta_{0}$ is the static dielectric constant . The direct application of this potential in Eq.~\eqref{U} leads
to a logarithmic divergence, and therefore, a screened Coulomb potential has to be taken into account.
According to the B-H approach, the potential can be written in a more rigorous form as 
$\phi(r) = \pm e/{\zeta{_0} r} \left(exp\left(-{r}/{r_0}\right)\right)$, 
where $r_0$ is the radius of ion field screening given by the following equation \cite{Askerov2010}, 
\begin{equation}
	\label{r0static}
	r_{0}^{-2}(n,k) = \frac{4\pi{e^2}}{\zeta_{0}}\int{-\frac{\partial f^{(0)}(\epsilon,\mu,T)}{\partial \epsilon_{n,k}}}g(\epsilon)d\epsilon~, 
\end{equation}
where $g(\epsilon)$ is the density of states, which in the calculations
is obtained numerically on an energy grid
with spacing $d\epsilon$ sampled over $N_k$ ${\bf{k}}$-points
\begin{equation}
	\label{dos2}
	g(\epsilon) = \int \sum_n \delta(\epsilon - \epsilon_{n,{\bf{k}}}) \frac{d{\bf{k}}}{8\pi^3} = \frac{1}{\Omega N_k}\sum_{n,{\bf{k}}} \frac{\delta(\epsilon - \epsilon_{n,{\bf{k}}})}{d\epsilon}~.
\end{equation}
From Eqs.~(\ref{tau3}),~(\ref{1PT}), and~(\ref{U}), the RT for the scattering of
charge carriers by ionized defects can be written for each band $n$ as 
\begin{equation}
\label{tau_def}
\tau_{def}(n,k) = \frac{\hbar\zeta{_0}{^2}}{2{\pi}{e^4}{N_i}F_{def}(k)}k^2 \left\vert\frac{\partial \epsilon_{n,k}}{\partial k}\right\vert 
\end{equation}
where
\begin{equation}
F_{def}(k) = ln(1+\eta) - \frac{\eta}{1+\eta}~, 
\end{equation}
is the screening function, with $\eta = (2kr_0)^2$.

Since crystalline As$_{2}$Se$_{3}$ is anisotropic, Eqs.~\eqref{r0static} and \eqref{tau_def} should be considered for each crystalline direction separately, according to Eqs. 21 and 22 in Ref.~\citenum{Chaves2021}. From the direction dependent relaxation times, one can obtain the corresponding TDKs and from them the direction dependent transport coefficient tensors can be determined. 


The calculation of carrier transport properties
were carried out using an in-house modified version of the \texttt{BoltzTraP} code~\cite{Madsen2006}.
The \texttt{BoltzTraP} code solves the linearized BTE using a Fourier interpolation 
of the band structure computed within the density functional theory (DFT) framework. 
The code computes all the integrals required to calculate the thermoelectric transport properties.

\subsubsection{DFT Calculation of Crystalline Structure and Band Structure}

Crystalline structure optimization and band structure calculations were performed using
DFT within the generalized gradient approximation (GGA) 
using the formulation of Perdew-Burke-Ernzerhof 
(PBE)~\cite{Perdew1996}, as implemented in the VASP package~\cite{Kresse1993,Kresse1996}.
To model the ion cores, projector augmented-wave (PAW)
pseudopotentials~\cite{Kresse1999} were used.
The electronic wave functions were expanded in a plane-wave basis-set
with a kinetic-energy cutoff of \num{700} eV. To sample the BZ,
a 5 x 7 x 15 Monkhorst-Pack ${\bf{k}}$-point 
mesh was used for the force on the ions calculations. 
Subsequently, to determine the transport properties, we used a finer ${\bf{k}}$-mesh 
of 15 x 19 x 45, which ensures the calculated transport coefficients to be converged. 
Within \texttt{BoltzTraP} the original ${\bf{k}}$-mesh is then interpolated
onto a mesh five times as dense ($M/N_{KS} = 5$).
We used experimental structure~\cite{Stergiou1985} as starting configuration and
relaxed the lattice parameters and atomic positions until
all atomic force components were smaller than $\num{1.0e-5}$ eV/$\AA$.

The weak interactions between layers along the $b$ direction were
accurately described by incorporating van der Waals (vdW) corrections to DFT, 
according to the D3 approach as given by Grimme \textit{et al.}~\cite{Grimme2010}.
We found that DFT-D3 performed reasonably well in predicting both the lattice
parameters and the internal atomic positions of As$_2$Se$_3$. 
We found that $a = 12.216\:\AA$, $b = 10.039\:\AA$ and $c = 4.259\:\AA$, 
deviating by $1.1$\%, $1.3$\% and $0.58$\%, respectively, from the experimental lattice 
parameters~\cite{Stergiou1985}. In particular, the calculated angle between the a- and c-axes is 90.526$\degree$, 
which in very good agreement with the experiment, 90.458$\degree$ \cite{Stergiou1985}.

Our calculations indicate that the material has an indirect band gap of \num{1.21} eV, which is in agreement with previous calculations \cite{Antonelli1986}. It is well known that DFT calculations underestimates the band gap when compared with the experimental one, which in this case is \num{2.0} eV \cite{Althaus1980}. In our calculations of the thermoelectric transport properties, the experimental band gap is used. As it is usually done, this is accomplished by shifting upward rigidly the calculated conduction bands \cite{Chen2016,Li2019}.

\subsubsection{Energetics and Concentration of Defects}


In order to estimate the equilibrium concentration of antisites in As$_2$Se$_3$, we first proceed to determine the formation energy
of a single antisite defect, i.e., an As atom at a Se site and a Se atom at an As site, through total energy calculations. 
A pair of atoms is placed at the swapped positions as far as possible from each other inside the supercell.
Since the number of atoms of each species remains the same upon the formation of the defect, the formation energy of a single antisite is calculated by simply taking
the difference between the total energy of a supercell containing a defect and the total energy of a supercell of a perfect crystal.
We obtained a formation energy, $\varepsilon_{def}$, for an antisite  of 1.91 eV, which is in very good agreement with previous total energy calculations 
that yielded 1.88 eV \cite{Tarnow1989}.

\begin{figure}
	\centering
	\includegraphics[width=0.50\textwidth]{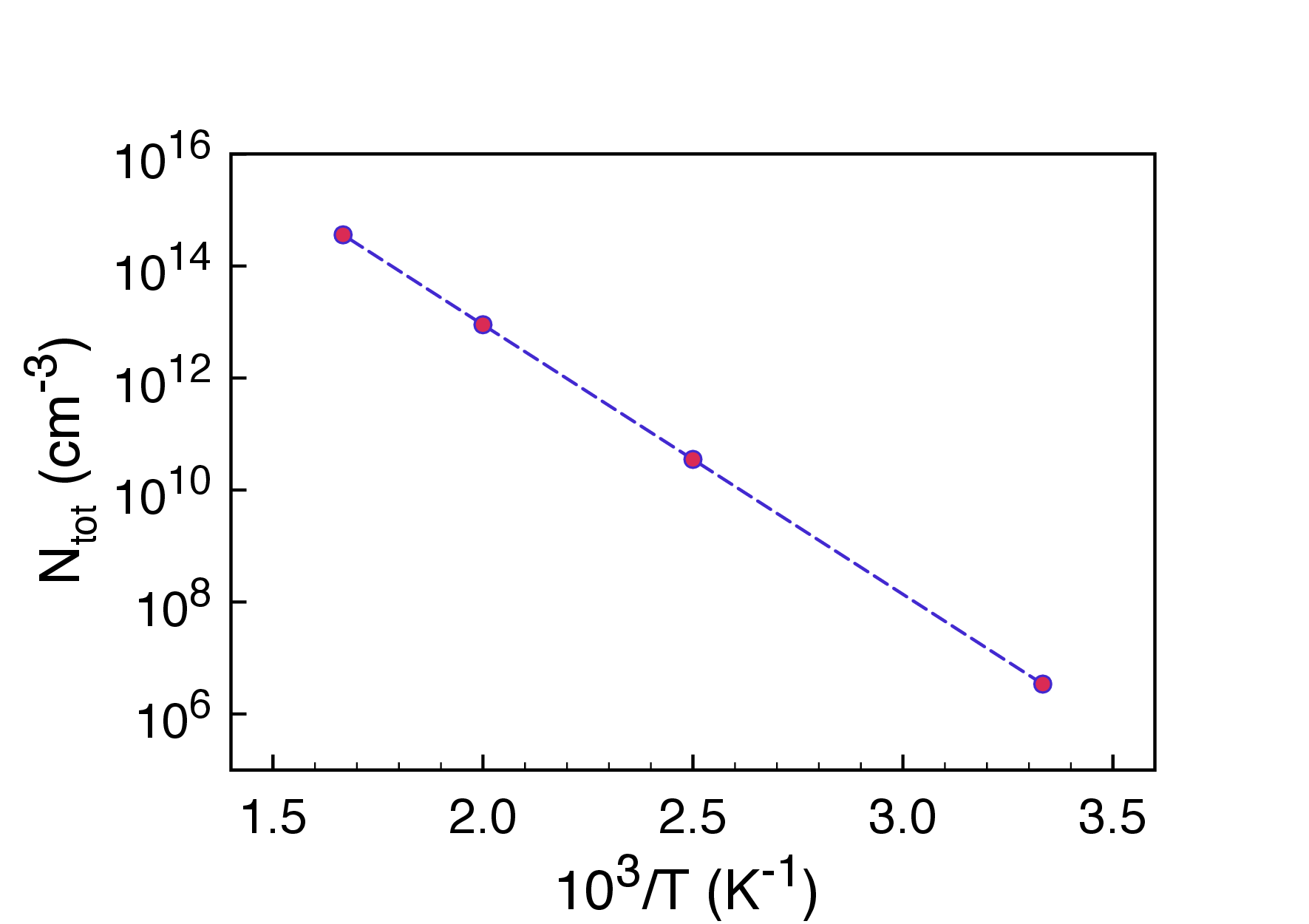}
	\vspace{-3mm}
	\caption{Equilibrium concentration of charged defects as a function of inverse temperature. The symbols correspond to results at 300~K, 400~K, 500~K, and 600~K, respectively.} 
	\label{equilibrium}
\end{figure} 

The estimation of the equilibrium concentration of antisites requires a more elaborate calculation, which will be summarized as follows. 
First, we have to compute the number of ways, $\Omega_{def}$, one can form N$_{def}$ pairs of atoms if there are N$_{As}$ arsenic atoms
and N$_{Se}$ selenium atoms per unit of volume. The total number of pairs As-Se that can be formed is given by the product of the
number of ways, $\Omega_{As}$, one can pick N$_{def}$ arsenic atoms among N$_{As}$ atoms and the number of ways, $\Omega_{Se}$, 
one can pick N$_{def}$ selenium atoms among N$_{Se}$ atoms, $\Omega_{def} = \Omega_{As}\cdot\Omega_{Se}$. This result stems from the fact that
the order in which the atoms are swapped is irrelevant, since atoms of the same species are indistinguishable. From $\Omega_{def}$, 
one can calculate the configurational entropy of the system, $S_{conf}=k_B ln(\Omega_{def})$.
Assuming that N$_{def}$ $\ll$ N$_{As}$ and N$_{def}$ $\ll$ N$_{Se}$ one can consider that the change in the vibrational entropy upon the formation of the defects
is negligible. Thus, the configurational entropy accounts for the whole change in the entropy of the crystal due to the formation of the
defects. Note that here we are taking the perfect crystal as the reference for both internal energy and entropy.
Neglecting changes in the volume, one can write from the first law of thermodynamics 
\begin{equation}
\frac{1}{T} = \frac{\partial S}{\partial E}, 
\end{equation}
where $E = N_{def}\varepsilon_{def}$. Therefore, we have
\begin{equation}
\label{1stlaw}
\frac{1}{T} = \frac{k_B}{\varepsilon_{def}}\frac{\partial ln(\Omega_{def})}{\partial N_{def}}.
\end{equation}
From Eq. \ref{1stlaw}, one can obtain the equilibrium concentration of antisite defects
\begin{equation}
\label{concentration}
N_{def}^{eq} = (N_{As}N_{Se})^{1/2} e^{-\frac{\varepsilon_{def}}{2k_BT}}.
\end{equation}

Therefore, the total concentration of charged defects in equilibrium ($As^+$ and $Se^-$) is given by $N_{tot}=2 N_{def}^{eq}$. 
To our knowledge, Eq. \ref{concentration} has previously appeared in Ref. \citenum{Brunst1985}, however, details of its derivation cannot be found, either there or elsewhere.

The concentrations, $N_{As} = 1.53 \times 10^{-22}$ cm$^{-3}$ and $N_{Se} = 2,30 \times 10^{-22}$ cm$^{-3}$ are obtained considering that in the 20-atom unit cell, whose volume is 5.22 $\times 10^{-22}$ cm$^{-3}$,  there are 8 As atoms and 12 Se atoms. Fig. \ref{equilibrium} depicts the results from Eq. \ref{concentration} in the temperature range from 300~K to 600~K. Due to the relatively large formation energy, the equilibrium concentration of charged defects is low, of the order of 10$^{14}$ cm$^{-3}$ at 600~K.

\section{Results and Discussion}

\begin{figure}
	\centering
	\includegraphics[angle=-90,width=0.50\textwidth]{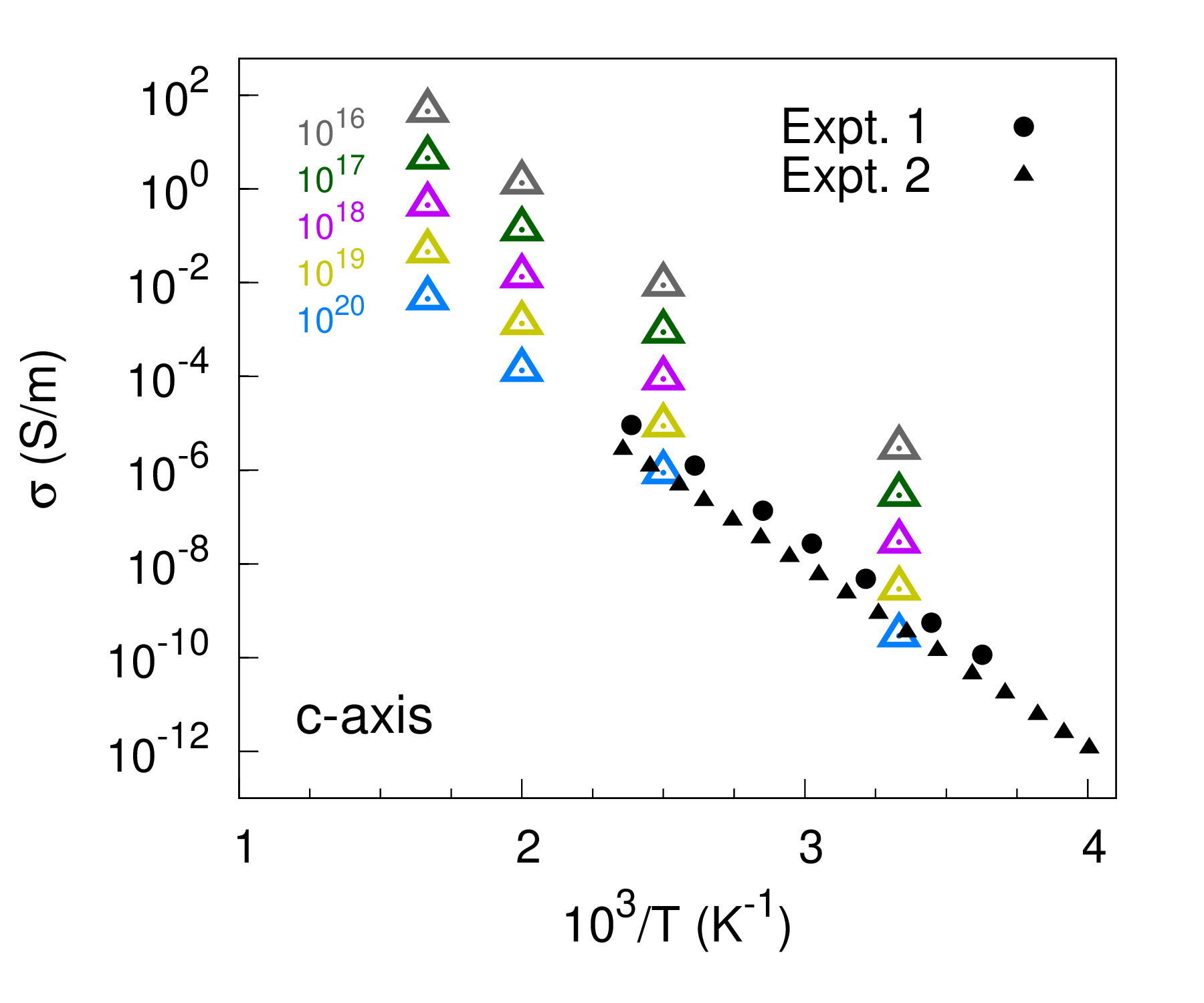}
	\vspace{-3mm}
	\caption{Electrical conductivity as a function of inverse temperature for defects concentrations ranging from 10$^{16}$cm$^{-3}$ to 10$^{20}$cm$^{-3}$. Experimental results from two different samples are from Ref. \citenum{Brunst1985}.} 
	\label{conductivity}
\end{figure} 

In order to determine the thermoelectric transport properties, two quantities should be specified beforehand, namely, the chemical potential or Fermi level and the concentration of charged defects. For the former, we used the experimental chemical potential provided in Ref. \citenum{Brunst1985}. Regarding the latter, we performed calculations for different values of the concentration of ionized defects. Our results for the electrical conductivity as a function of temperature, together with experimental ones from Ref. \citenum{Brunst1985}, are shown in Fig. \ref{conductivity}. Our findings indicate good agreement with the experimental results for concentrations between 10$^{19}$ cm$^{-3}$ and 10$^{20}$ cm$^{-3}$ .

Next, we compare these results with those for the equilibrium concentration of charged defects determined using Eq. \ref{concentration}. As it has been previously pointed out, even at the temperature of 600~K, the equilibrium concentration of charged defects is of only 3.6 $\times$  $10^{14}$ cm$^{-3}$, which is orders of magnitude smaller than that yielded by conductivity calculations. This large discrepancy strongly suggests that the large concentration of defects in the samples investigated experimentally is out-of-equilibrium, possibly formed during crystal growth due to kinetic factors and/or deviations from stoichiometry.
A similar out-of-equilibrium concentration of defects has been proposed by Zhou \emph{et al.} \cite{Zhou2018} to explain the high carrier density in SnSe. This last result is quite interesting, because it indicates that, in principle, there is plenty of room for the reduction of the concentration of antisite defects in the material. That could lead to Fermi level depinning and allow to utilize the material in thermoelectric applications. 

\begin{figure}
	\centering
	\includegraphics[angle=-90,width=0.50\textwidth]{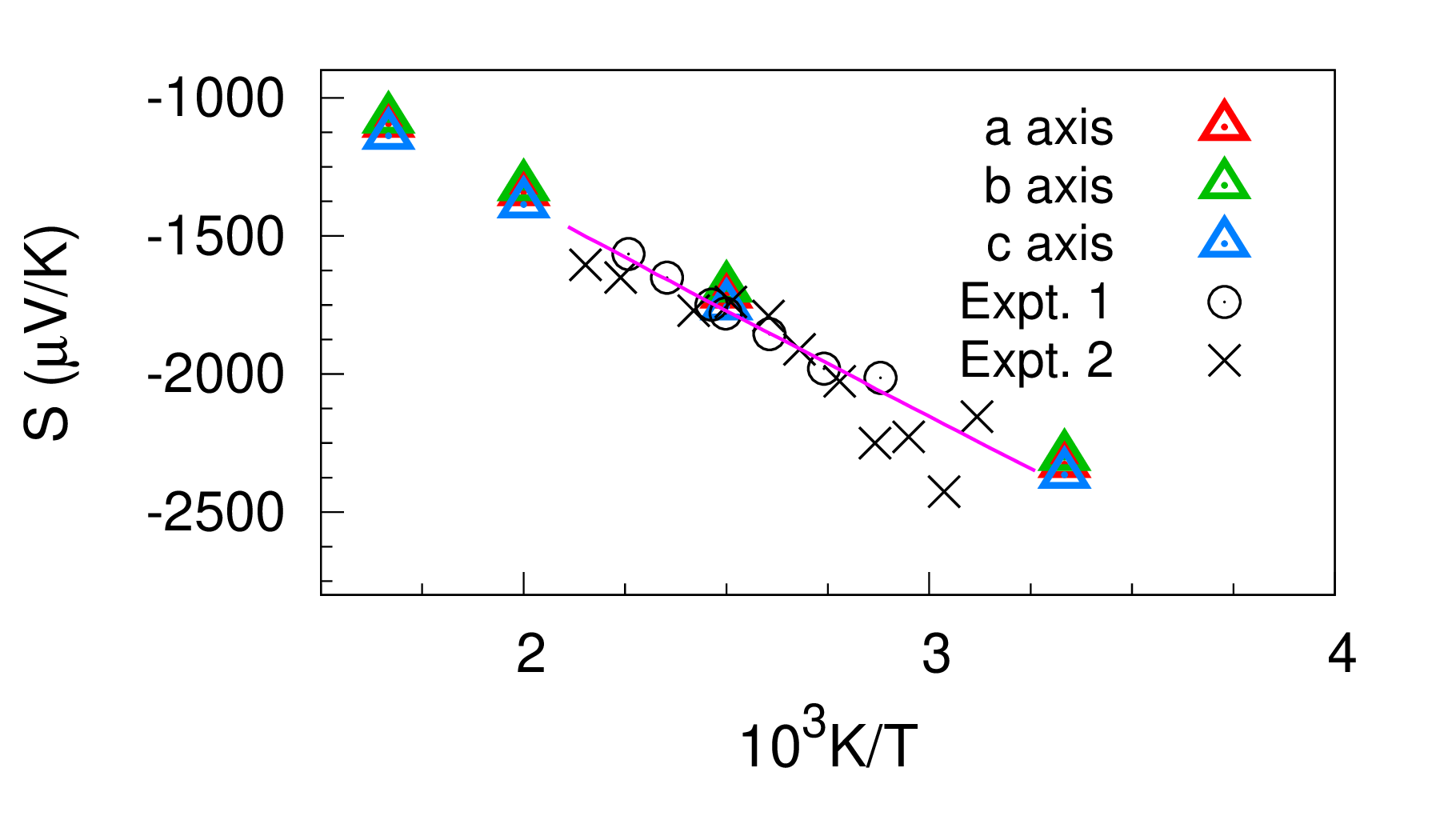}
	\vspace{-3mm}
	\caption{Seebeck coefficient as a function of inverse temperature for defect concentration of 10$^{20}$cm$^{-3}$. Experimental results from two different samples are from Ref. \citenum{Brunst1985}.} 
	\label{Seebeck}
\end{figure}

In Fig. \ref{Seebeck} we show the results for the Seebeck coefficient calculated assuming the concentration of charged defects of $10^{20}$ cm$^{-3}$. In this case also, we can see a good agreement between our results and those from Ref. \citenum{Brunst1985}. 

Let us now assume that the concentration of defects can be reduced and consequently Fermi level depinning can be achieved, what thermolectric properties the material would exhibit? Aside from the ability to dope the material, which would provide relevant increase in the carrier density, the reduction in charged defects concentration would decrease substantially scattering rates. Our calculations indicate that at the highest temperature considered of 600~K, if the concentration of charged defects could be decreased from $\sim$ 10$^{20}$ cm$^{-3}$ to $\sim$ 10$^{17}$ cm$^{-3}$, the carrier relaxation time along the c-direction due to scattering by defects would increase from a value $\sim$ 0.01 fs to $\sim$ 10 fs. These values are similar to relaxation times found in materials such as SnSe \cite{Chaves2021}. Since a full first-principles calculations of the carrier-phonon scattering is beyond the scope of this work, because of the large size of the unit cell, we adopt the approximation that the full scattering relaxation time is independent of the wavevector, $\bf{k}$, of the charge carrier. In this case, the relaxation time, $\tau_{n,k}(\mu,T) = \tau(\mu,T)$, can be taken out of the integral in $d\bf{k}$ in Eq. \ref{TDK} in the calculation of the TDKs. In order to check the validity of this approximation, we utilize the following procedure, since in the formula to compute the Seebeck coefficient the relaxation time cancels out, we are able to vary the chemical potential in order to adjust the Seebeck coefficient to the experimental value \cite{Brunst1985} for a set of values of $T$. Our results for the chemical potential obtained from this procedure are in good agreement with the experimental findings from Ref. \citenum{Brunst1985} indicating that indeed the approximation is reliable in this case. 

\begin{figure}
	\centering
	\includegraphics[width=0.50\textwidth]{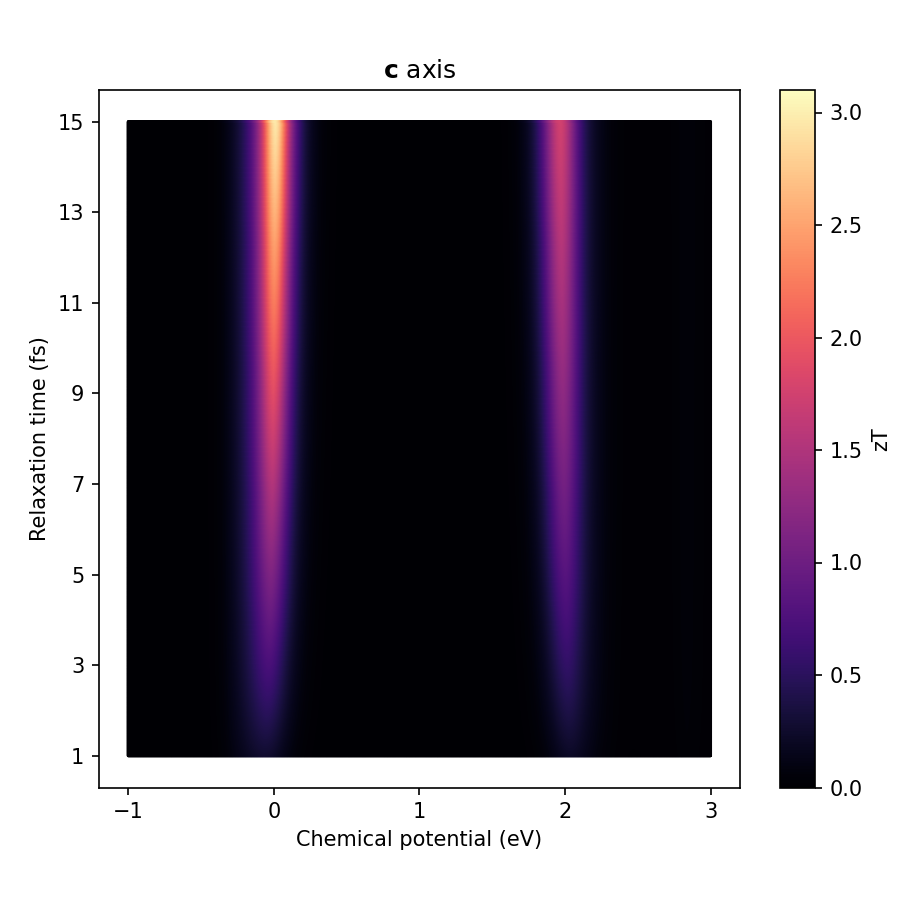}
    \vspace{-3mm}
	\caption{Figure of merit of As$_2$Se$_2$ as a function of the Fermi level and relaxation time. }
	\label{zT}
\end{figure} 

Fig. \ref{zT} displays the figure of merit $zT$ along the c-axis as a function of the chemical potential and relaxation time. The thermal conductivity used in the calculations is given by the sum of two contributions from carriers and from phonons. For the latter contribution, we employed that obtained in Ref. \citenum{Gonzalez-Romero2018}. For the p-type material, $zT$ can reach the high value of 3, while for the n-type case $zT$ is about 2, in both cases for relaxation times above 10 fs. 

It should be emphasized that the calculations in Ref. \citenum{Gonzalez-Romero2018} considered only the perfect crystal case. The inclusion of the effect of defects would further decrease the lattice contribution to the thermal conductivity, therefore, increasing $zT$.

\subsection{Summary}

Based on the solution of the linearized BTE within the RTA and using 
smooth Fourier interpolation of the KS eigenvalues and their corresponding velocities,  
we calculated the thermoelectric transport properties of As$_2$Se$_3$. The calculations were done using a
methodology that incorporates anisotropy in the determination of the RTs \cite{Chaves2021}. 
Comparison of our results with experimental findings \cite{Brunst1985} indicate that scattering by charged defects 
dominates the relaxation processes in this compound.
According to our calculations, the concentration of charged defects is about 10$^{19}$ cm$^{-3}$. Previous
DFT-based calculations \cite{Tarnow1989} indicate that antisites, are such defects, namely, threefold coordinated Se$^+$ ions at As sites
and twofold coordinated As$^-$ ions at Se sites. 
We have also estimated the equilibrium concentration of these defects, which, due to their relatively high formation energy,
is of only 10$^{14}$ cm$^{-3}$ at the highest temperature considered of 600~K. 
These results suggest that the large concentration of defects in the samples investigated experimentally
is the result of kinetic factors and/or deviations from stoichiometry. 
Therefore, at least in principle, there are no fundamental reasons to prevent a reduction of the charged defects. 
Based on this conclusion, we computed the thermoelectric transport properties assuming that the
concentration of charged defects could be reduced, which would increase the RTs to values similar to those
found in other chalcogenide compounds such as SnSe \cite{Chaves2021}. For relaxation times above 10 fs, we obtain
for the p-type material $zT$ as high as 3, while for the n-type case $zT$ of about 2. Those remarkable values of $zT$ are somehow underestimated,
since they were obtained using the phonon thermal conductivity previously calculated by Gonzalez-Romero \emph{et al.} \cite{Gonzalez-Romero2018},
which did not consider phonon scattering by defects.
 
%

\vspace{0.5 cm}

\section*{Acknowledgement}

ASC and AA gratefully acknowledge support from the Brazilian agencies CNPq and FAPESP under 
Grants \#2010/16970-0, \#2013/08293-7, \#2015/26434-2, \#2016/23891-6, \#2017/26105-4, and \#2019/26088-8. 
MAS gratefully acknowledges support from the Brazilian Agencies CAPES and CNPq.
The calculations were performed at CCJDR-IFGW-UNICAMP in Brazil.

\bibliographystyle{apsrev4-1}


\end{document}